\documentclass[aps,pre,twocolumn,superscriptaddress,floatfix]{revtex4}
\usepackage{graphicx,color}
\usepackage[dvips,unicode,colorlinks,linkcolor=blue,citecolor=blue,urlcolor=blue]{hyperref}
\usepackage{times}

\begin{document}
\title{Thermal conductivity of the chain with an asymmetric pair interaction}
\author{A. V. Savin}
\email[]{asavin@center.chph.ras.ru}
\affiliation{Semenov Institute of Chemical Physics, Russian Academy of Sciences,
Moscow 119991, Russia}

\author{Yuriy A. Kosevich}
\email[]{yukosevich@gmail.com, yuriy.kosevich@ecp.fr}
\affiliation{Semenov Institute of Chemical Physics, Russian Academy of Sciences,
Moscow 119991, Russia}
\affiliation{Laboratoire d'Energ\'etique Mol\'eculaire et Macroscopique, CNRS UPR 288, 
Ecole Centrale Paris, Grande Voie des Vignes, 92295 Ch\^atenay-Malabry, France}

\date{\today}

\begin{abstract}
We provide molecular dynamics simulation of heat transport in one-dimensional molecular chains
with different interparticle pair potentials. We show that the thermal conductivity is finite
in the thermodynamic limit in the chains with the potential, which allows for bond dissociation.
The Lennard-Jones, Morse and Coulomb potentials belong to such type of potentials.
The convergence of the thermal conductivity is provided by phonon scattering on the locally
stretched loose interatomic bonds at low temperature and by the many-particle scattering
at high temperature. On the other hand, the chains with the confining pair potential, which
does not allow for the bond dissociation, posses anomalous (diverging with the chain length)
thermal conductivity. We emphasize that the chains with the symmetric or asymmetric
Fermi-Pasta-Ulam potential or with the combined potentials containing parabolic
confining potential all exhibit anomalous heat transport.

\end{abstract}

\maketitle

\section{Introduction}

The heat conductivity of low-dimensional system have attracted intensive studies \cite{LLP03}.
The main problem is to derive from first principles, on the atomic level, the thermal conductivity
(TC) $\kappa$ as the coefficient, in Fourier law, between the heat flux and temperature gradient,
$J=-\kappa \nabla T$.
In one-dimensional (1D) case the Fourier law for the chain of $N$ molecules is reduced to the following:
\begin{equation}
J=\kappa (T_+-T_-)/(N-1)a, \label{f1}
\end{equation}
where $T_\pm$ is temperatures of left and right chain ends, $a$ is a lattice period.
For small temperature difference, $\Delta T=(T_+-T_-)\ll T=(T_++T_-)/2$, Eq. (\ref{f1}) allows to find
the dependence of thermal conductivity on the length  $L=(N-1)a$ and temperature of the chain $T$:
\begin{equation}
\kappa(N,T)=J(N-1)a/(T\delta T),~\delta=(T_+-T_-)/T\ll 1. \label{f2}
\end{equation}
The Fourier law  (\ref{f1}) is fulfilled if the following finite limit exists:
$$
\bar\kappa(T)=\lim_{N\rightarrow\infty}\kappa(N,T).
$$
The chain has a finite TC in this case, while it has the anomalous (diverging with the chain length)
TC in the case of  $\kappa\rightarrow\infty$ for $N\rightarrow\infty$.

To date, there are numerous works devoted to the numerical modeling of heat transfer in 1D lattices.
Anomalies of heat transport in 1D nonlinear systems are well known since the time of the
famous work of Fermi, Pasta, Ulam \cite{FPU}. In integrable systems (harmonic grid, Toda chain,
chain of rigid disks) the heat flux $J$ does not depend on the chain length $L$,
therefore the thermal conductivity diverges: $ \kappa (L)\sim L$ for $ L \rightarrow \infty $.
Here, the heat transfer is carried out by non-interacting quasiparticles, so the energy is not
dissipated during the heat transfer. Non-integrability of the system is a necessary but not a
sufficient condition for the existence of the normal thermal conductivity.
On the examples of a chain with symmetric potential of the Fermi-Pasta-Ulam (FPU) \cite{LRP,LLP1,LLP2},
disordered harmonic chain \cite{RG,CL,D}, diatomic 1D gas of colliding particles \cite{D1,STZ02,GNY},
and the diatomic Toda lattice \cite{H}, it was shown that some non-integrable systems can also
have infinite (diverging with the system size) thermal conductivity. Here, the thermal conductivity
increases as a power function of the length: $\kappa\sim L^\alpha$, for $L\rightarrow \infty$,
with $0 <\alpha <1$.

On the other hand, the chain with the on-site potential can have finite thermal conductivity.
The convergence of the thermal conductivity with the system size was shown for the Frenkel-Kontorova
chain \cite{HLZ98, SG03}, for the chain with the sinh-Gordon on-site potential \cite{TBSZ},
for the chain with $\phi^4$ on-site potential \cite{HLZ00,AK00} and for the chain of hard disks
with substrate potential \cite{GS04}. The essential feature of these models is the presence of an
external potential, which models the interaction of the chain with the substrate. These systems do
not possess the translational invariance and the total momentum is not conserved therein.
It has been suggested in \cite{H} that the presence of an external potential plays a key
role for the convergence of thermal conductivity in the system. It was conjectured an infinite
thermal conductivity for all isolated one-dimensional lattices, where the absence of external
potential leads to the conservation of the system total momentum. This hypothesis was refuted
in the papers \cite{giardina, savin2}, in which it was shown that the isolated chain of coupled
rotators (a chain with a periodic interparticle potential) has a finite thermal conductivity.

Anomalous thermal conductivity of isolated chains is connected with a weak scattering of
long-wavelength phonons having long mean free paths. A mechanism of effective scattering of
long-wavelength phonons should exist for the occurrence of finite thermal conductivity.
For the rotator chain \cite{savin2}, such a mechanism is the scattering of phonons by roto-breathers.
In recent papers \cite{ZZWZ12,CZWZ12} the thermal conductivity of an isolated chain with
asymmetric pair inter-particle potential was studied. It is claimed in these papers that with a
certain degree of interaction asymmetry, the thermal conductivity measured in nonequilibrium
conditions converges in the thermodynamical limit. The authors of \cite{ZZWZ12,CZWZ12}
attribute the convergence of the thermal conductivity to the uneven thermal expansion of the
asymmetric chain. Without a doubt, the thermal expansion of the chain itself can not be the
reason for the convergence of thermal conductivity, because such expansion is present in the
chain with an asymmetric Toda potential, which is an integrable system!
Modeling of heat transfer in \cite{ZZWZ12} was performed with the help of nonequilibrium
MD simulation with the Nose-Hoover heat baths. The deterministic Nose-Hoover thermostat
is not intended to simulate heat transfer. Its use leads to modeling of the dynamic system which
consists of two attractors connected by a one-dimensional chain \cite{FHLZ}
and may lead to incorrect results \cite{LLM}. Also in the works \cite{ZZWZ12,CZWZ12}
the effective mechanism of the scattering of long-wave acoustic phonons was not found.
Dispel the convergence of long-wavelength phonons providing heat. All this has made
necessary to conduct a more detailed modeling of the heat transport in 1D systems.

The aim of this work is to simulate the heat transport in molecular chains with asymmetric
potentials of the pair inter-particle interaction in the framework of 1) nonequilibrium
molecular dynamics using Langevin thermostat and 2) equilibrium dynamics using the Green-Kubo approach.
It will be shown that the chains with unbounded asymmetric potentials of the pair inter-particle
interaction can have finite thermal conductivity. For instance the chains with the asymmetric pair
potentials which allow for bond dissociation (similar to the Lennard-Jones and Morse potentials)
possess  finite thermal conductivity in the thermodynamic limit. A chain with a purely repulsive
interaction potentials, like the chain with Coulomb interaction between nearest neighbors, also
has final thermal conductivity. Thermal conductivity convergence in such chains is
caused by strong (anomalous) Rayleigh scattering of long-wave phonons by the fluctuations of local
extension at low temperatures, and by many body scattering at high temperatures.
On the other hand, the presence of the inter-particle pair interaction which limits such fluctuations
and prevents the transformation of the 1D lattice into a 1D gas of colliding
particles, results in anomalous (diverging with the system size) thermal conductivity of the 1D chain.
Hence the chains with the FPU potential, either symmetric or asymmetric,  and the chains with any
combined pair potential which includes the parabolic confining potential, have infinite thermal
conductivity in the thermodynamic limit. The reason is that such chains do not allow for the bond
dissociation or rupture. From our studies we can conclude that the thermal transport of a one-dimensional
chain will be anomalous break if the asymmetric potential of the inter-particle pair interaction
grows not slower than the square of the relative distance.

\section{Model}

We consider the molecular chain consisting $N$ units. In a dimensionless form, the Hamiltonian of
the chain can be written as
\begin{equation}
H=\sum_{n=1}^N\frac12\dot{u}_n^2+\sum_{n=1}^{N-1}V(u_{n+1}-u_{n}-r_0),
\label{f3}
\end{equation}
where $ u_n $ -- dimensionless coordinate of $n$-th node, the dot denotes differentiation
with respect to the dimensionless time $t$, $V(r)$ -- dimensionless interaction potential between
nearest neighbors normalized terms of $V(0)=0$, $V'(0)=0$, $V''(0)=1$, $r_0$ -- equilibrium bond length.
From now on we assume $r_0=1$.

To simulate the heat transfer in the chain we will use the stochastic Langevin thermostat.
Consider a finite chain of $N_++N+N_-$ links. If the interaction potential $V(r)$
does not suppose possibility of rupture of the bond, then we take the chain with the free ends,
and if it admits --- chain with fixed ends. Put the $N_+$ right boundary nodes in the Langevin
thermostat with temperature $T_+$, and $N_-$ nodes in the left-hand edge thermostat
Langevin with a temperature of $T_-$. The corresponding system of equations of motion of the chain
is given by
\begin{eqnarray}
\ddot{u}_n&=&-\partial H/\partial u_n -\gamma\dot{u}_n+\xi_n^+,~~
n\le N_+,\nonumber\\
\ddot{u}_n&=&-\partial H/\partial u_n,~~n=N_++1,...,N_++N,\label{f4}\\
\ddot{u}_n&=&-\partial H/\partial u_n -\gamma\dot{u}_n+\xi_n^-,~~
n\ge N_++N+1, \nonumber
\end{eqnarray}
where $\gamma=0.1$ is relaxation coefficient of the particle velocity, $\xi_n^\pm $ simulates the interaction
with thermostat white Gaussian noise normalized by the conditions
\begin{eqnarray*}
\langle\xi_n^\pm (t)\rangle=0,~~
\langle\xi_n^+(t_1)\xi^-_k(t_2)\rangle=0, \\
\langle\xi_n^\pm (t_1)\xi_k^\pm (t_2)\rangle = 2\gamma T_\pm\delta_{nk}\delta(t_2-t_1).
\end{eqnarray*}

The system of equations of motion (\ref{f4}) integrated numerically. After the onset of thermal equilibrium
chain with thermostats and education stationary heat flow was distributed along the chain
temperature $T_n = \langle\dot{u}_n^2\rangle_t $ and the local heat flux
$$
J_n=\bar{a}\langle j_n \rangle_t,~~\mbox{where}~~j_n=-\dot{u}_nV '(u_{n}-u_{n-1}-r_0),
$$
$\bar{a}$ -- average value of the bond length at $T=(T_++ T_-)/2$ (because of thermal expansion at $T>0$,
the average bond length exceeds the equilibrium bond length at $T=0$: $\bar{a}\ge r_0=1$).
The following values were used in the numerical simulation:
$T_\pm=(1\pm 0.1)T$, $\gamma=0.1$, $N_\pm=40$, $N=20$, 40, 80, ..., 20480.

\begin{figure}[tb]
\includegraphics[angle=0, width=1\linewidth]{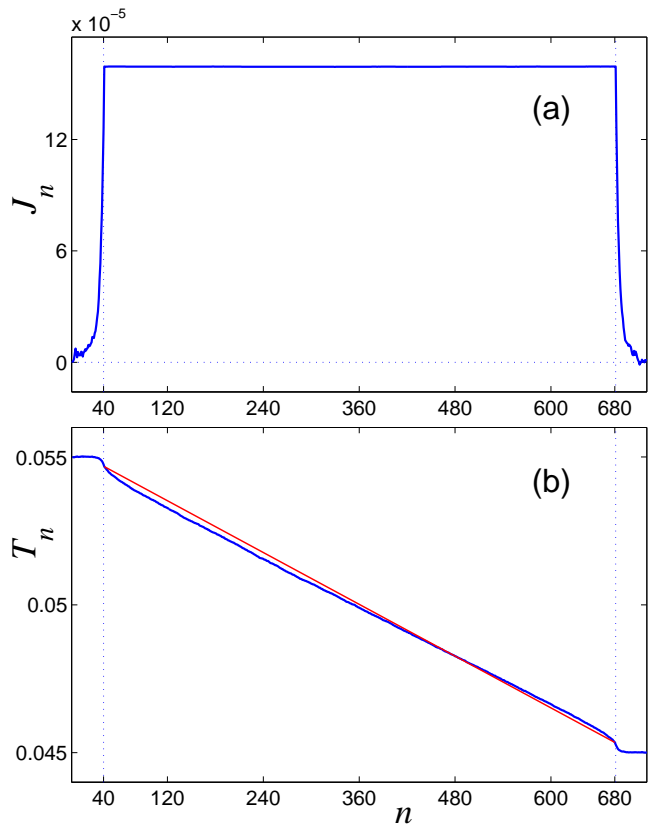}
\caption{
Distribution of (a) local heat flux $J_n$ and (b) local temperature $T_n$ in
FPU chain (parameters $\alpha=2$, $\beta=1$) for $N_\pm=40$, $N=640$, $T_+=0.055 $, $T_-=0.045$.
The straight line shows the linear temperature gradient which is used in obtaining
thermal conductivity $\kappa(N)$.
}
\label{fig01}
\end{figure}

This method of thermalization overcomes the problem of the thermal boundary resistance.
The distribution of the local energy flux $J_n$ and temperature profile $T_n$ along the chain
are shown in Figure \ref{fig01}. In the steady-state regime, the heat flux through each link
at the central part of the chain should remain the same, i.e., $J_n\equiv J$, $N_+<n\le N_++N$.
This property can be employed as a criterion for the accuracy of numerical modeling and can also
be used to determine the characteristic time for achieving the steady-state regime and calculation of
$J_n$ and $T_n$. Figure \ref{fig01} suggests that the flux is constant along the
central part of the chain thus suggesting we achieved the required regime.

At the central part of the chain, we observe the almost linear gradient of the temperature distribution,
so that we can define the coefficient of thermal conductivity as
\begin{equation}
\kappa(N)=J(N-1)\bar{a}/(T_{N_++1}-T_{N_++N}).
\label{f5}
\end{equation}

Thermal conductivity can also be found through the Green-Kubo formula \cite{GK}
\begin{equation}
\kappa_c=\lim_{t\rightarrow\infty}\lim_{N\rightarrow\infty}\frac{1}{NT^2}\int_0^tc(\tau)d\tau,
\label{f6}
\end{equation}
where the flow-stream function correlations
$c(t)=\langle J_s(\tau)J_s(\tau-t)\rangle_\tau $,
and the total heat flux in the chain $J_s(t)=\bar{a}^2\sum_n j_n (t)$.

To find the correlation function $c(t)$, we consider a finite cyclic chain with $N=10^4$
units, fully immersed in the Langevin thermostat with temperature $T$. After the onset of the
thermal equilibrium with the thermostat, we disable chain interaction with the thermostat
and model further on the dynamics of an isolated thermalized chain.
To increase the accuracy of the correlation function measurement,  we average the found value over
$10^4$ independent realizations of the initial thermalization of the chain.

Here the question of the convergence of the heat conductivity is reduced to the question
of the rate of decay of the correlation function $c(\tau)$ for $\tau\rightarrow\infty$.
The chain has a finite conductivity if the rate of decrease is sufficient for the convergence
of the integral (\ref{f6}) and the anomalous thermal conductivity otherwise.

\section{Fermi-Pasta-Ulam Potential}

\begin{figure}[tb]
\includegraphics[angle=0, width=1\linewidth]{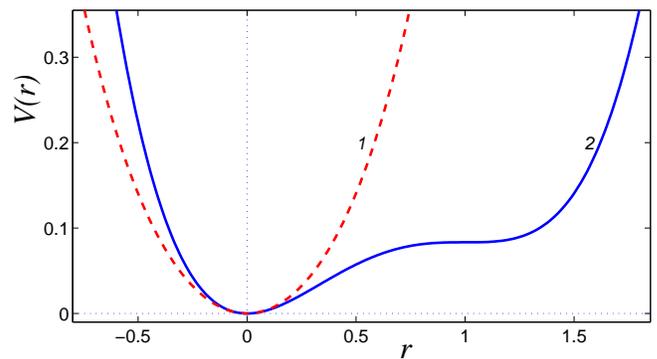}
\caption{
The form of the symmetric ($\alpha=0$, $\beta=1$, curve 1) and asymmetric ($\alpha=2$, $\beta=1$
curve 2) FPU potential (\ref{f7}).}
\label{fig02}
\end{figure}

We consider here a chain with $\alpha$-$\beta$-FPU pair potential:
\begin{equation}
V(r)=r^2/2-\alpha r^3/3+\beta r^4/4.
\label{f7}
\end{equation}
When $\alpha>0$, the FPU potential function is asymmetric -- see Fig. \ref{fig02}.
For definiteness, we take the following values of the parameters: $\alpha =2$, $\beta=1$

\begin{figure}[tb]
\includegraphics[angle=0, width=1\linewidth]{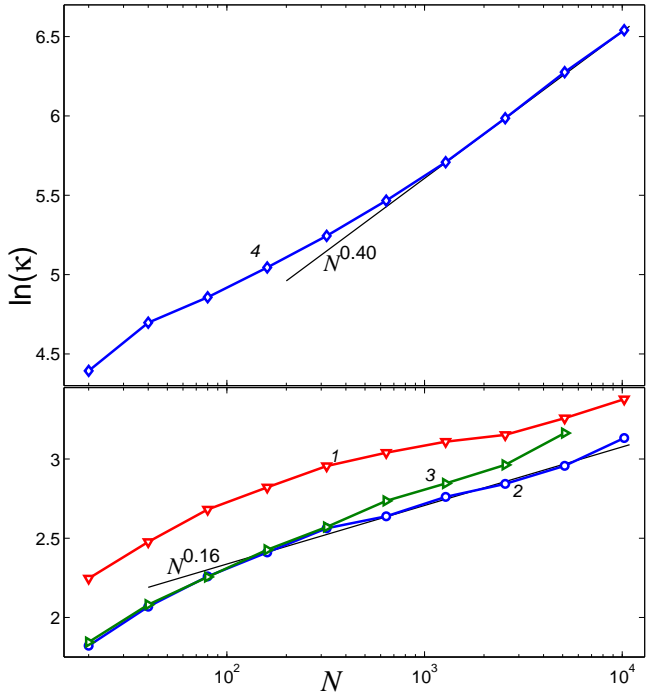}
\caption{
The dependence of the natural logarithm of the thermal conductivity $\kappa$ on the
length $N$ of the internal part of the chain with an asymmetric FPU potential (\ref{f7}) with
$\alpha=2$, $\beta=1$ and temperature $T=0.025$ (curve 1), $T=0.05$ (curve 2), $T=0.1$ (curve 3),
and for the chain with a symmetric FPU potential ($\alpha=0$, $\beta=1$) at $T=0.05$ (curve 4).
Straight solid lines show the power-law dependencies $ N^{0.16}$ and $N^{0.40}$.
}
\label{fig03}
\end{figure}

The FPU potential does not allow for molecular chain breaking, therefore we consider the chain
with free ends. The dependence of the thermal conductivity $\kappa$ on the length of the
internal part $N$ is shown in Fig. \ref{fig03}. As can be seen from this figure, the asymmetry
of the FPU potential does not result in the convergence of the thermal conductivity, it only
decreases the divergence rate. Thus, at temperature $T=0.05$ in the chain with the symmetric FPU
potential (for $\alpha=0$) the thermal conductivity increases with the length as $N^{0.40}$,
while in the chain with the asymmetric FPU potential (with $\alpha=2$) it increases as $N^{0.16}$.
This divergence reduction is caused by the fact that in the chain with asymmetric FPU pair
potential a new channel of phonon scattering opens -- the scattering of phonons by strongly
stretched loose bonds.

\begin{figure}[tb]
\includegraphics[angle=0, width=1\linewidth]{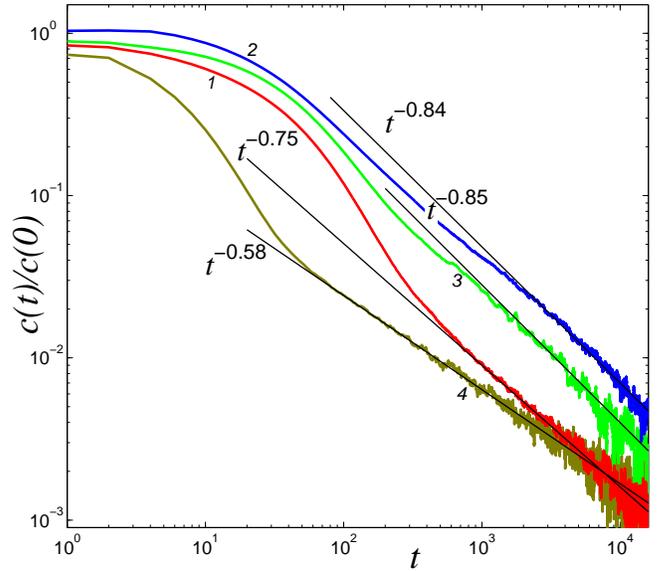}
\caption{
The power law decay of the correlation function $c(t)$ for the chain with an asymmetric FPU
potential (\ref{f7}) ($\alpha=2$, $\beta=1$, temperature $T=0.05$, curve 1) and for the chain
with hyperbolic potential (\ref{f11}) at $T=1$: $\delta=0$, $b=0$ (curve 2);
$\delta=1$, $b=0$ (curve 3); $\delta=0$, $b=1$, $c=4$ (curve 4).
The straight lines give the power-law dependencies
$t^{-0.58}$, $t^{-0.75}$, $t^{-0.84}$, and $t^{-0.85}$.
}
\label{fig04}
\end{figure}

Analysis of the behavior of the correlation function $c(t)$ for $t\rightarrow\infty$ confirms
the divergence of the thermal conductivity in the FPU chain. At $t\rightarrow\infty$ correlation
function $c(t)$ decreases as a power function $t^{-\delta}$ with the exponent $\delta<1$,
see Fig.~\ref{fig04}, curve 1. Therefore the integral in the Green-Kubo formula (\ref{f6}) diverges.

\section{Lennard-Jones and Morse Potentials}

\begin{figure}[tb]
\includegraphics[angle=0, width=1\linewidth]{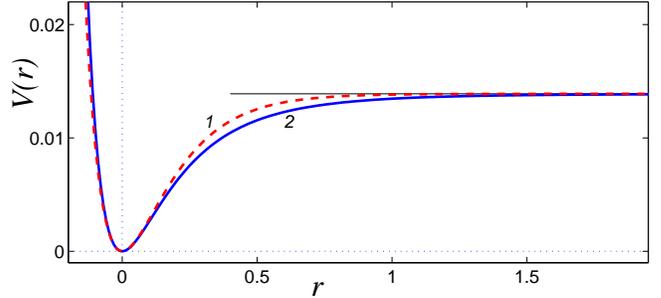}
\caption{
The form of the Morse (\ref{f9}) and the Lennard-Jones (\ref{f8}) potentials, curves 1 and 2 respectively.
Straight solid line sets the binding energy of $\varepsilon = 1/72$.
}
\label{fig05}
\end{figure}

We use the following form for the Lennard-Jones pair potential,
\begin{equation}
V(r)=4\varepsilon\{[\sigma/(1+r)]^6-1/2\}^2,
\label{f8}
\end{equation}
with the parameter $\sigma=2^{-1/6}$ and binding energy $\varepsilon=1/72$,
and for the Morse pair potential:
\begin{equation}
V(r)=\varepsilon[\exp(-\beta r)-1]^2,
\label{f9}
\end{equation}
with the parameter $\beta=1/\sqrt{2\varepsilon}=6$.
For these potentials $V(0)=0$, $V'(0)=0$, $V''(0)=1$, $\lim_{r\rightarrow +\infty}V(r)=\varepsilon$.
The characteristic form of these two potentials are shown in Fig.~\ref{fig05}.

Since the Lennard-Jones (\ref{f8}) and Morse (\ref{f9}) pair potentials have finite
binding energies, the chains with these pair potentials and free edges are not stable with respect
to thermal fluctuations. After some time, such chains will necessarily break and heat flux
over them will stop. Therefore the thermal conductivity can be modeled only for such chains
with fixed ends. For this purpose, the equations of motion (\ref{f4}) should be imposed by the
following boundary conditions:
\begin{equation}
u_1\equiv 0,~~~u_{N_++N+N_-}\equiv(N_++N+N_--1)a,
\label{f10}
\end{equation}
where the parameter $a\ge r_0$ characterizes the density of the chain $d=r_0/a$. In these chains
strong density fluctuations can occur, so we will take into account the interaction of $n$-th particle
with the left edge thermostat only if its coordinate satisfies $u_n(t)<N_+a$, and the interaction
with the right thermostat only if it satisfies $u_n(t)>(N_++N)a$.
\begin{figure}[tb]
\includegraphics[angle=0, width=1\linewidth]{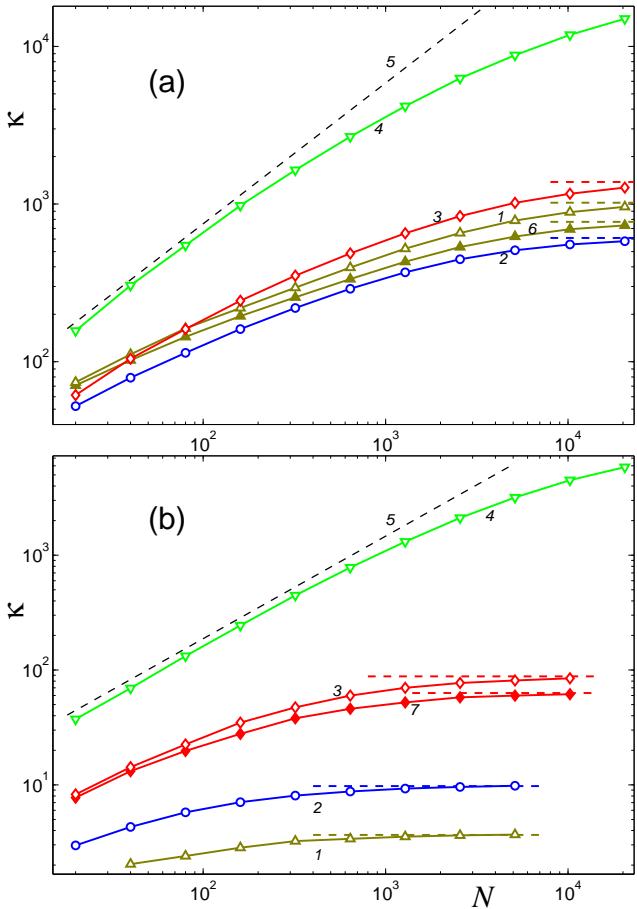}
\caption{
The dependence of the thermal conductivity $\kappa$ on the length $N$ of the internal part of the chain
with Lennard-Jones pair potential for chain density (a) $d=1$ and (b) $d=2/3$. Dependencies are
shown at $T=0.002$, 0.005, 0.02, 0.2 (curves 1, 2, 3, 4). Straight line 5 shows the power law $N^{0.89}$.
Curves 6 and 7 give the dependencies for the chain with the Morse pair potential at $T=0.002$ and
$T=0.02$. The horizontal straight lines give the values of the thermal conductivity obtained with the
use of the Green-Kubo formula.
}
\label{fig06}
\end{figure}
\begin{figure}[tb]
\includegraphics[angle=0, width=1\linewidth]{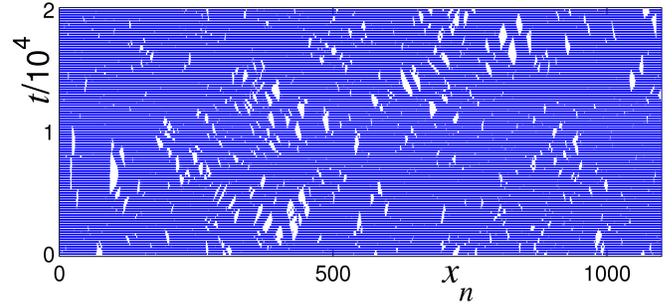}
\caption{
Time dependence of the distribution of particles in the chain with Lennard-Jones potential with
length $N=1000$. The density of the chain is $d=1/1.1$, temperature $T=0.002$.
Each particle is represented as an interval of unit length with the center at $x_n$.
Vertical white areas show the loose stretched bonds and the time of life of such fluctuations
correspond to the height of the area.
}
\label{fig07}
\end{figure}

The dependence of the thermal conductivity of the chain with the Lennard-Jones pair potential
on its length is shown in Fig.~\ref{fig06}. We can conclude from the results of the numerical
simulation that the chain with the density of $d=2/3$ at low temperatures has a finite thermal conduction.
The sequence $\kappa(N)$ converges at $T=0.002$ and $T=0.005$. The trend of the convergence is
observed at high temperatures as well, but the convergence is slower and is not fully realized
for the used lengths $N\le 10240$. Equilibrium modeling confirms the convergence of thermal conductivity.
Here, the correlation function $c(t)$ tends to zero exponentially, so the integral in the
Green-Kubo formula always converges. Both methods give the same limiting values of thermal
conductivity, see Fig.~\ref{fig06}(b).

For low temperatures, where the thermal transport is provided by phonons, the convergence of
the thermal conductivity is caused by Rayleigh phonon scattering at the fluctuations of the
local bond stretching (bond dissociation). For the particle density $d<1$, the loose stretched
bonds are present in the lattice and their time of life increases with temperature decrease.
As one can see in Fig.~\ref{fig07},  there are 3 or 4 fluctuations of strong bond stretching
with time of life $t_b>10^3$ in the chain with $N=1000$ particles at temperature $T=0.002$
and particle density $d=1/1.1$. In such a system the phonons will have a finite mean free path,
from one loose stretched bond to another.
\begin{figure}[tb]
\includegraphics[angle=0, width=1\linewidth]{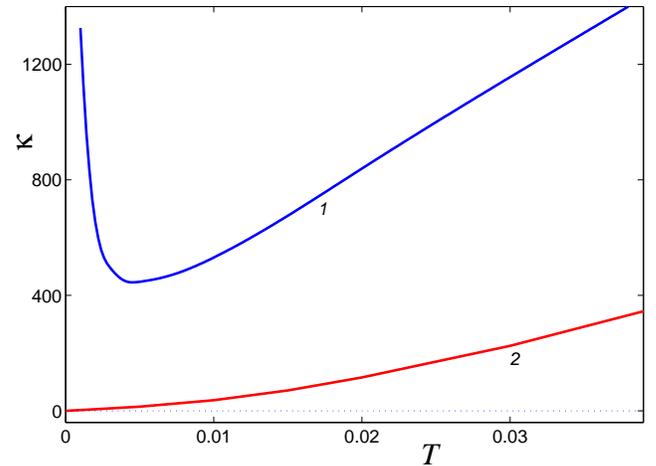}
\caption{
Temperature dependence of the thermal conductivity $\kappa$ of the Lennard-Jones chain with
length $N=2560$ for chain density $d=1$ (curve 1) and $d=2/3$ (curve 2).
}
\label{fig08}
\end{figure}

The phonon thermal conductivity shows the tendency of the convergence in the thermodynamic
limit in the chain with particle density $d=1$.
The convergence is slower in such chain than that in the chain with the density $d<1$. In the
chain with the nominal density $d=1$, the probability of fluctuation of strong bond
stretching is lower than in the dilute chain with density $d<1$.

Temperature dependence of the thermal conductivity $\kappa$ is shown in Fig.~\ref{fig08}.
At normal density $d=1$, thermal conductivity is nonmonotonic function of temperature.
Thermal conductivity increases both at $T\rightarrow 0$ and $T\rightarrow\infty$. This is due
to the fact that at low temperatures the chain behaves almost like a harmonic chain, which has
an infinite thermal conductivity, while at high temperatures it behaves as a gas of rigid disks,
which is also characterized by the anomalous thermal transport.

But the situation is changed in the diluted chain with density $d=2/3$. Here the thermal
conductivity increases monotonously with temperature.
For $T\rightarrow\infty$, the thermal conductivity goes to infinity because the system
behaves as a gas of rigid disks. But the thermal conductivity goes to zero for $T\rightarrow 0$
in such system. It is related with the property of the dilute chain  with density $d<1$
that such chain has fluctuations of strong bond stretching which time of life goes to infinity
for $T\rightarrow 0$. The phonon transport becomes impossible in this case and the
thermal conductivity goes to zero for $T\rightarrow 0$ in such a system.

Note that the performed in Ref. \cite{chen04} modeling of the thermal conductivity of
Ar nanowires has shown that the quasi-three-dimensional rod made from the particles
interacting through the Lennard-Jones potential has a finite (normal) thermal conductivity.
This shows that the convergence of the thermal conductivity, with the system length, is faster
in quasi-three-dimensional rod than in one-dimensional chain.

In the chain with Morse pair potential, thermal conductivity shows the same behavior as
in the chain of Lennard-Jones, see Fig.~\ref{fig06}. Therefore, all the results obtained
for the Lennard-Jones chain are also valid for the Morse chain.

\section{Hyperbolic potentials}

\begin{figure}[tb]
\includegraphics[angle=0, width=1\linewidth]{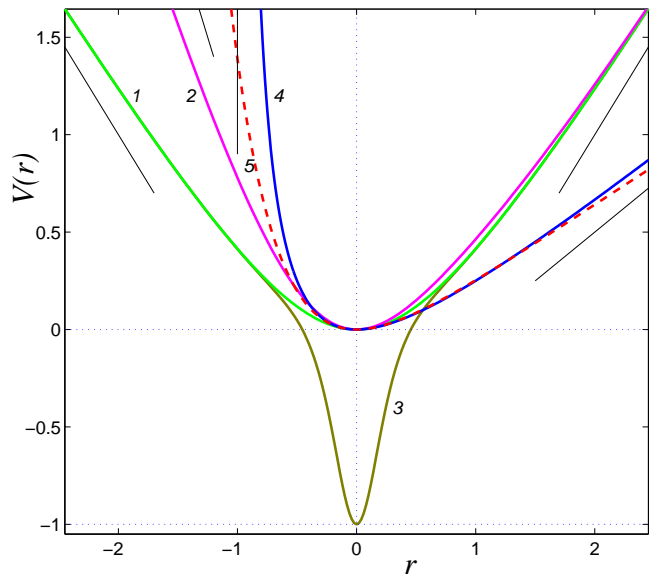}
\caption{
View of hyperbolic potential (\ref{f11}) with a parameters: $\delta=0$, $b=0$ (curve 1);
$\delta=1$, $b=0$ (curve 2); $\delta=0$, $b=1$, $c=4$ (curve 3)
and an asymmetric potential with a hard core (\ref{f12}) (line 4).
Straight lines show the potential asymptotes. The dashed line 5 shows the Toda potential (\ref{f13})
with the parameter $b=2.5$.
}
\label{fig09}
\end{figure}

\begin{figure}[tb]
\includegraphics[angle=0, width=1\linewidth]{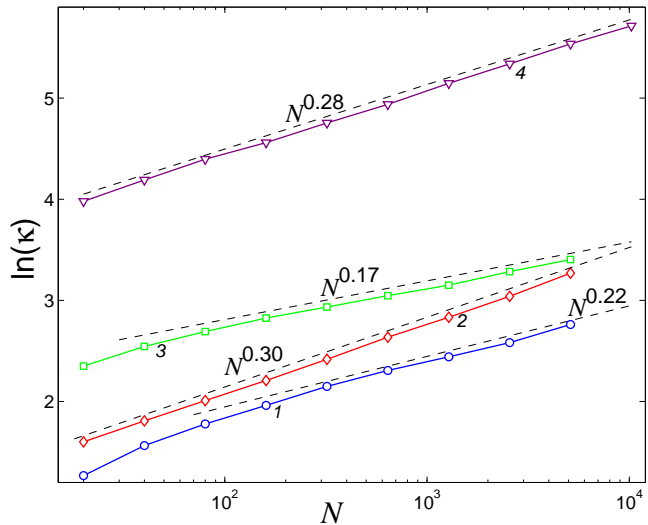}
\caption{
Dependence of the coefficient of thermal conductivity $\kappa$ on the length $N$ of the central
part of two-mass Toda chain (masses $m_1=1$, $m_2=2$, ...) at $T=0.2$ (curve 1),
and of the chain with the hyperbolic potential (\ref{f11}) at $T=1$ with the parameters: $\delta=0$,
$b=0$ (curve 2); $\delta=1$, $b=0$ (curve 3); $\delta=0$, $b=1$, $c=4$ (curve 4).
The straight dashed lines give the power-law dependencies $N^{0.22}$, $N^{0.30}$, $N^{0.17}$ and $N^{0.28}$.
}
\label{fig10}
\end{figure}

We consider the interaction potentials having the form of a hyperbola with linear asymptotes
at $r\rightarrow\pm\infty$:
\begin{equation}
V(r)=[1+\frac12\delta(1-\tanh r)](\sqrt{1+r^2}-1)-\frac{b}{\cosh^2(cx)},
\label{f11}
\end{equation}
where the parameter $\delta\ge 0$ characterizes the asymmetry of the potential.
For $r\rightarrow +\infty$ and $r\rightarrow -\infty$, the hyperbolic potential (\ref{f11}) has the
asymptotes $r-1$ and $(1+\delta)(r +1)$, respectively. Thus for $\delta=0$ or $\delta=1$ and $b=0$,
the potential (\ref{f11}) has the form of a symmetric or asymmetric hyperbole (see Fig.~\ref{fig09},
curves 1 and 2).
The following  potential also has the form of asymmetric hyperbole:
\begin{equation}
V(r)=r^2/2(r+1),~~~r>-1.
\label{f12}
\end{equation}
The potential (\ref{f12}) describes the interaction with the hard core.
It is defined only for the relative displacements of $r>-1$.
For $r\rightarrow -1$, the interaction energy tends to infinity
[energy grows like $1/2(r +1)$], and for $ r\rightarrow\infty $
potential increases as a linear function $(r-1)/2$ (see Fig.~\ref{fig09}, curve 4).
In contrast to the symmetric potential FPU, which is characterized by hard anharmonicity, the
hyperbolic potential (\ref{f11}) has a soft anharmonicity.
Our modeling shows that thermal transport in the chain with the potential (\ref{f11})
is anomalous. For the symmetric potential (parameters $\delta=0$, $b=0$), at
temperature $T=1$ the thermal conductivity $\kappa(N)$ grows as $N^{0.30}$
(see Fig.~\ref{fig10}, curve 2), and the correlation function $c(t)$ decays as $t^{-0.84}$
(see Fig.~\ref{fig04}, curve 2).
The asymmetry of the potential reduces the rate of divergence.
For the asymmetric potential (parameters $\delta=1$, $b=0$), the thermal conductivity $\kappa(N)$
grows as $N^{0.17}$ (see Fig.~\ref{fig10}, curve 3), and the correlation function $c(t)$
decays as $t^{-0.85}$ (see Fig.~\ref{fig04}, curve 3).
For the symmetric potential with additional well (parameters $\delta=0$, $b=1$, $c=4$), the thermal
conductivity $\kappa(N)$ grows as $N^{0.28}$ (see Fig.~\ref{fig10}, curve 4),
and the correlation function $c(t)$ decays as $t^{-0.58}$ (see Fig.~\ref{fig04}, curve 4).
\begin{figure}[tb]
\includegraphics[angle=0, width=1\linewidth]{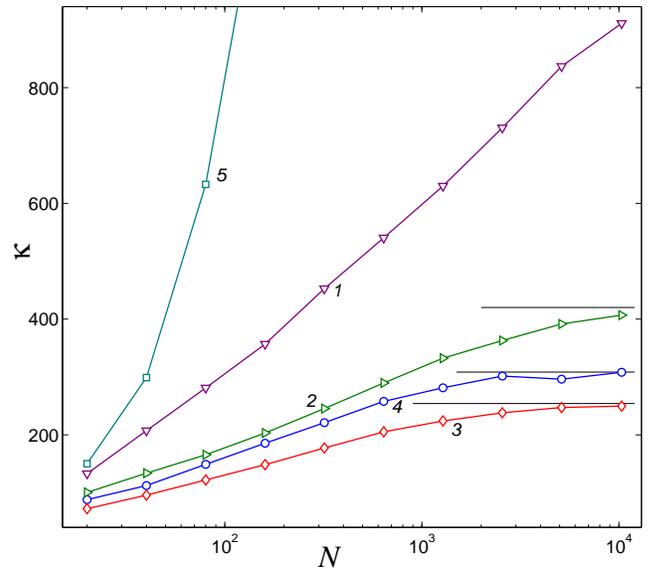}
\caption{
Dependence of the coefficient of thermal conductivity $\kappa$ on the length $N$ of the central
part of the chain with the hard core potential (\ref{f12}) at
$T=0.05$, 0.1, 0.2, 0.5 (curves 1, 2, 3, 4) and chain with Toda potential (\ref{f13}) at $T=0.2$
(curve 5). The straight lines give the values of the thermal conductivity obtained with the use of the
Green-Kubo formula.
}
\label{fig11}
\end{figure}

\begin{figure}[tb]
\includegraphics[angle=0, width=1\linewidth]{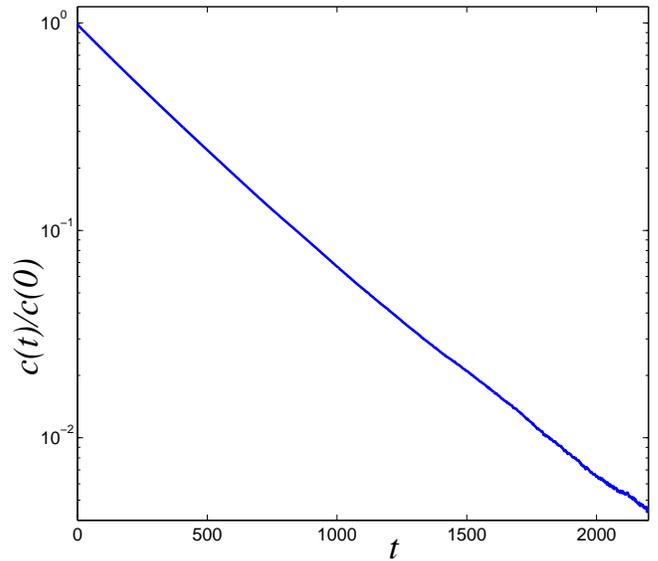}
\caption{
Exponential decrease of the correlation function $c(t)$ for the chain with asymmetric hyperbolic
potential (\ref{f12}) at $T=0.2$.
}
\label{fig12}
\end{figure}

In Fig.~\ref{fig11} we show the dependence of the thermal conductivity $\kappa$ on the length
of the central part $N$ of the chain with asymmetric potential with a hard core (\ref{f12}).
As one can see, at low temperatures $T=0.05$ thermal conductivity first increases $\kappa\propto\log(N)$,
then the growth rate starts to slow down. At high temperatures, the thermal conductivity is convergent.
This is also confirmed by the behavior of the correlation functions. At $t\rightarrow\infty$,
correlation function decays exponentially, $c(t)\propto\exp(-\lambda t) $, see Fig.~\ref{fig12}.
Therefore, the Green-Kubo formula (\ref{f6}) gives a finite value of thermal conductivity.
Moreover, both approaches (equilibrium and non-equilibrium MD modeling) give
the same limiting values for $N\rightarrow\infty$, see Fig.~\ref{fig11}, which confirms the correctness
of the numerical simulation.

\section{Toda potential}

For comparison, we also consider the Toda pair potential:
\begin{equation}
V(r)=b^{-2}[\exp(-br)+br-1],
\label{f13}
\end{equation}
where the parameter $b>0$. To be definite, we take the value of the parameter $b=2.5$ at which
at low temperatures Toda potential (\ref{f13}) is well approximated by the
asymmetric potential (\ref{f12}), see Fig.~\ref{fig09}, curves 4 and 5.

The chain with the Toda pair potential (\ref{f13}) (Toda lattice) is a completely integrable system.
The MD simulation shows that the values of the boundary temperatures $T_{N_++1}$, $T_{N_++N}$
and the value of the heat flux $J$ do not depend on the length of the chain central part $N$
(there is no energy dissipation  because the heat transport is performed by non-interacting
with each other nonlinear excitations of the Toda lattice). Thus, in view of (\ref{f5}) the
thermal conductivity $\kappa(N)$ grows as $N$, see Fig.~\ref{fig11}.

Toda lattice eases to be completely integrable if its atoms have different masses.
The simplest case is the two-mass lattice when the odd sites have the dimensionless mass
$m_1=1$ while the even sites have mass $m_2=m>1$. Modeling of the thermal conductivity of such a chain
was carried out in \cite{H}, where it was shown that for $m=2$ the thermal conductivity $\kappa$
diverges as $ N^{0.35} $. Our MD simulations show a slower divergence: $\kappa(N)\sim N^{0.22}$
for $ N\rightarrow\infty$, see Fig.~\ref{fig10}, curve 1.
Therefore the isotopic disorder in the lattice does not lead to the convergence of thermal conductivity.
The convergence of the thermal conductivity in one-dimensional lattice can only be provided by the
non-linearity of the (pair) inter-atomic interaction.

\section{Purely repulsive potentials}

Now we consider the thermal conductivity of the chains with purely repulsive potentials:
\begin{eqnarray}
V(\rho)&=&1/\rho, \label{f14} \\
V(\rho)&=&1/\rho^{12}, \label{f15} \\
V(\rho)&=& \exp[-b(\rho-1)^3], \label{f16}\\
V(\rho)&=& \exp[-b(\rho-1)] \label{f17}\\
V(\rho)&=& \exp[-2b(\rho-1)]+2\exp[-b(\rho-1)] \label{f17a}
\end{eqnarray}
where $\rho$ is a distance between the interacting particles, and the parameter $b>0$.
Purely repulsive is also the following short-range potential:
\begin{eqnarray}
V(\rho) &=& 0~~\mbox{for}~~\rho\ge 1,\nonumber\\
V(\rho) &=& \exp\{-b[(\rho-1)+\alpha/(\rho-1)^2]\}~\mbox{for}~\rho<1,~~~~\label{f18n}
\end{eqnarray}
where parameters $b>0$, $\alpha>0$.

In order to obtain a stable system, we fix the length of the chain and
consider the chain of $ N $ sites with fix ends:
\begin{equation}
u_1(t)\equiv 0,~~~u_N(t)\equiv (N-1)a, \label{f18}
\end{equation}
Then the ground state of the chain with a fixed density of $d=1/a$ will be the
homogeneous lattice with period $a$: $u_n^0=(n-1)a$, $n= 1$,2, ...,N, in which the pair
inter-particle potential is given by one of the repulsive potentials (\ref{f14})--(\ref{f18n}).

In a chain with a fixed density, the Coulomb repulsion potential (\ref{f14})
can be replaced by the asymmetrical hyperbolic potential:
\begin{equation}
V(\rho)=\frac{1}{\rho}+\frac{\rho}{a^2}-\frac{2}{a}=r^2/a^2(r+a), \label{f19}
\end{equation}
where the relative displacement $r=\rho-a$ is introduced. Replacement of the Coulomb potential (\ref{f14})
by the asymmetric hyperbolic potential (\ref{f19}) does not change the
system of equations of motion of a chain with fixed ends (\ref{f18}). Therefore
one should expect that the chain with a repulsive Coulomb potential has
finite thermal conductivity.
\begin{figure}[tb]
\includegraphics[angle=0, width=1\linewidth]{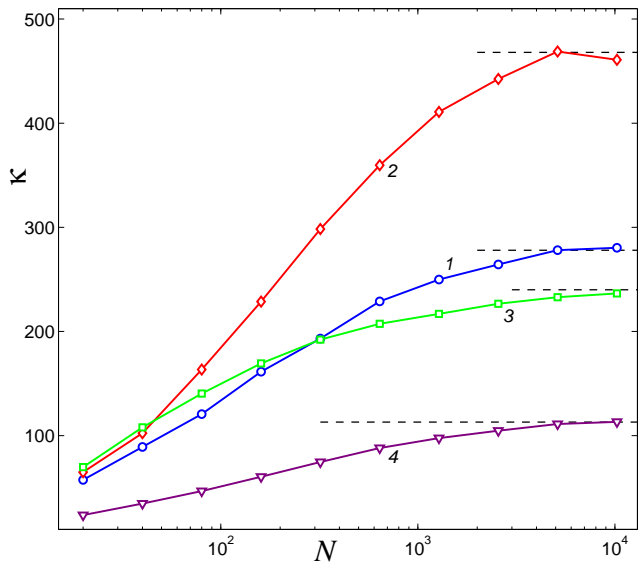}
\caption{
Thermal conductivity $\kappa$ versus length $N$ of the internal part of the chain
with a repulsive Coulomb potential (\ref{f14}) (line 1 for lattice period $a= 1$, line 2
for the period $a=2$), with short-range potential (\ref{f18n}) (line 3, for the period $a=3$,
parameters $b=2.5$, $\alpha=0.05$), and for a chain with a repulsive potential (\ref{f16})
(curve 4, for the period  $a=1$, parameter $b=2.5$). Temperature of the chain is $T=1$.
The horizontal straight lines give the values of the thermal conductivity obtained with the use of the
Green-Kubo formula.
}
\label{fig13}
\end{figure}

Direct modeling of heat transfer has shown that thermal conductivity of the chain with a
fixed density converges with the increase of the length, see Fig.~\ref{fig13}.
Analysis of the behavior of the correlation function also confirms the convergence of
thermal conductivity: at $t\rightarrow\infty$ the correlation function $c(t)$ tends to zero
exponentially. Both methods of non-equilibrium and equilibrium MD modeling  give the same limiting
value, for $N\rightarrow\infty$, of the thermal conductivity.
\begin{figure}[tb]
\includegraphics[angle=0, width=1\linewidth]{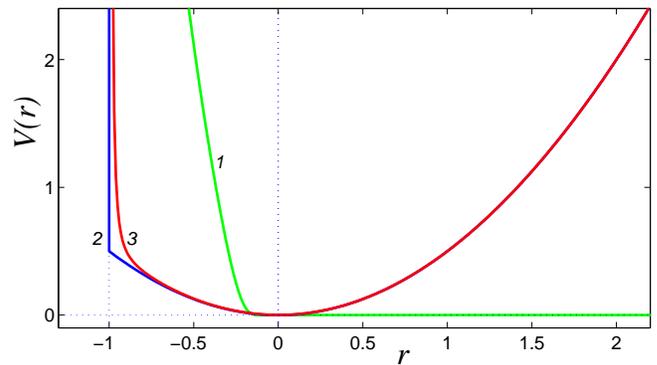}
\caption{
Form of the short-range potential (\ref{f18n}) (curve 1, the parameters $b=2.5$, $\alpha=0.05$),
vibro-impact potential (\ref{f25}) (curve 2), and  its smooth approximating
potential (\ref{f26}) (curve 3), where $r=\rho-1$ is the relative displacement.
}
\label{fig14}
\end{figure}

In order to understand the mechanism, which provides the convergence of the thermal conductivity, we
consider the chain with a repulsive short-range potential (\ref{f18n}). For definiteness we take
the parameter $b=2.5$, the parameter $\alpha=0.05$. The form of the potential is shown in Fig.~\ref{fig14},
line 1.

\begin{figure}[tb]
\includegraphics[angle=0, width=1\linewidth]{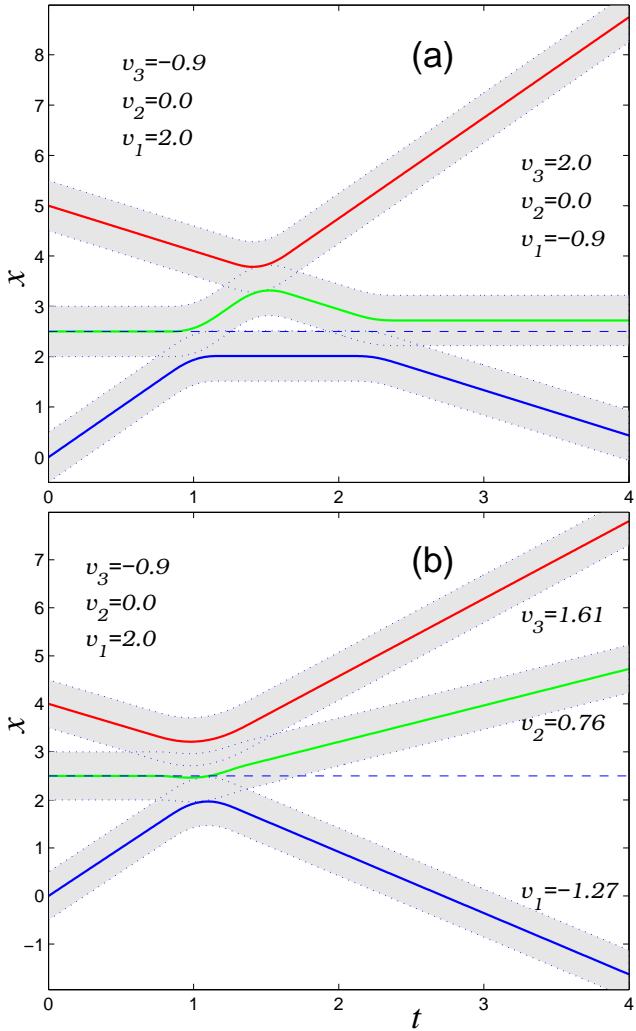}
\caption{
Three consecutive pair collisions (a) and one the three-body collision of three particles (b)
in a chain with a short-range potential (\ref{f18n}). The lines show the particle position $x$
versus time $ t $. Gray bars show the diameters of the particles $|x(t)-x|<0.5$ (particles interact
only when their diameters intersect). Particle velocities $v_1$, $v_2$, $v_3$ before and after
all the interactions are shown.
}
\label{fig15}
\end{figure}

The repulsion of the particles takes place only when the distance between them becomes $\rho<1$.
The collision of the two particles in such interaction will be elastic, but will occur
for a finite period of time, see Fig.~\ref{fig15}~(a). If all the interactions
between particles can be reduced to the two-body collisions only, the momentum would be
transmitted along the chain without scattering and the chain would have the infinite thermal conductivity.
That is the case of the chain consisting of stiff disks performing instantaneous elastic collisions.
If the collision does not occur instantaneously, but takes a finite interval of time, it makes possible
the many-body collisions, for example, the three-particle collision (collision of one particle
with a pair of interacting particles). 
An example of the three-particle collision is shown in Fig.~\ref{fig15}~(b).
As one can see, in the three-particle collision there is no complete transfer of momentum from one
extreme particle to another, some of the total energy remains at the central particle. Therefore
the many-body collisions in one-dimensional chain should lead to the scattering of energy and
finite thermal conductivity.

Modeling of the heat transfer along the chain of particles with the short-range pair interaction potential
(\ref{f18n}) shows the convergence of the thermal conductivity, see Fig.~\ref{fig13}.
Analysis of the behavior of the correlation function also confirms the convergence of the thermal
conductivity. Both methods of the non-equilibrium and equilibrium MD modeling give the same limiting
value, $N\rightarrow\infty$, of the thermal conductivity. From this we can conclude that the
convergence of the thermal conductivity in a lattice with the short-range repulsive potential
(\ref{f18n}) is provided by the many-particle collisions.

\begin{figure}[tb]
\includegraphics[angle=0, width=1\linewidth]{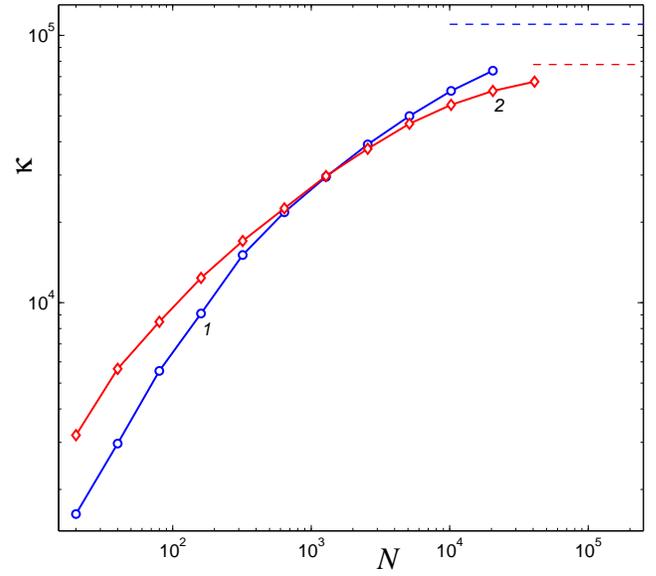}
\caption{
Thermal conductivity $\kappa$ versus length $N$ of the internal part of the chain with repulsive
potential (\ref{f15}) (curve 1, temperature $T=1$), and with potential (\ref{f17a}) (curve 2,
$b=6$, $T=10$). Chain density is $d=1$ (chain period  $a=1/d=1$).
The horizontal straight lines give the values of the thermal conductivity, which are obtained
with the use of the Green-Kubo formula.
}
\label{fig16}
\end{figure}

Chain with a repulsive potential (\ref{f15}) also has a finite thermal conductivity.
This potential leads to a more rigid collision than the Coulomb potential does.
With this potential, the time of pair collision is shorter and hence the effects caused by
the many-particle collisions should exhibit weaker, and the thermal conductivity
must converge slowly with the length increase.
The correlation function $c(t)$ decays in time exponentially.
At temperature $T=1$, the  Green-Kubo formula gives the finite  value of thermal conductivity,
$\kappa_c= 110000$. The dependence of the thermal conductivity $\kappa$ of the chain length $N$
is shown in Fig.~\ref{fig16}. As one can see, the convergence of the thermal conductivity should be
expected at lengths $N\sim 250000$.

  The chain with a repulsive potential (\ref{f16}) also has the final thermal conductivity.
Here, the correlation function $c(t)$ decays exponentially, and the Green-Kubo formula gives the limiting
value of $\kappa(N)$ for  $N\rightarrow\infty$, which coincides with the one given by
non-equilibrium simulations, see Fig.~\ref{fig13}.

In a chain with a fixed density, pure exponential repulsive potential (\ref{f17})
can be replaced by the Toda potential
\begin{eqnarray}
V(r)=e^{-br}+\beta e^{-ba}r-e^{-ba}[1+ba] \nonumber\\
=e^{-ba}[\exp(-br)+br -1], \label{f20}
\end{eqnarray}
where $r=\rho-a$ is the relative displacement. Replacement of the exponential potential (\ref{f17})
by the Toda potential (\ref{f20}) does not lead to a change in the equations of motion in the
chain with fixed ends (\ref{f18}). Therefore the chain with an exponential repulsive potential is
completely integrable system and has an infinite conductivity. In the direct simulation of the
heat transfer, the values of the boundary temperatures $T_{N_++1}$ and $T_{N_++N}$, and the value
of the heat flux $J$ do not depend on the length of the chain between the thermostats $N$ (the same
temperature is set throughout the whole central part of the chain).

Repulsive potential which is a sum of two exponential functions (\ref{f17a}) can not be replaced
by the Toda potential (\ref{f20}). The presence of the second exponential function in the
potential makes the molecular chain to be the non-integrable system. Our MD simulations show that
the chain with such repulsive potential has a finite thermal conductivity, and the correlation
function decays exponentially. For $b=6$ and temperature $T=10$, the  Green-Kubo formula gives the
value of the thermal conductivity $\kappa_c=77800$. The dependence of the thermal conductivity
$\kappa$ of the chain length $N$ is shown in Fig.~\ref{fig16}. As one can see, the convergence
of the thermal conductivity should be expected at chain length $N\sim 10^5$.

\section{Combined asymmetric potentials}

Now we consider the potentials which are the sum of the previously considered
asymmetric potentials with the harmonic (parabolic) potential. The addition of the parabolic
potential leads to the stabilization of the inter-atomic bonds. Below we study how this
stabilizing effects on the thermal conductivity of the chain.

\begin{figure}[tb]
\includegraphics[angle=0, width=1\linewidth]{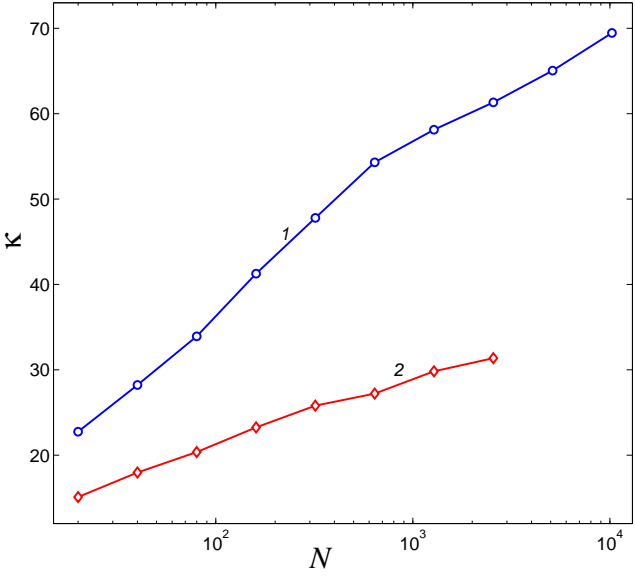}
\caption{
Dependence of the coefficient of thermal conductivity $\kappa$ on the length $N$ of the central
part of the chain with the combined potentials (\ref{f22}) (curve 1, $b=2.5$, $T=1$)
and (\ref{f23}) (curve 2, $T=0.5$).
}
\label{fig17}
\end{figure}

\begin{figure}[tb]
\includegraphics[angle=0, width=1\linewidth]{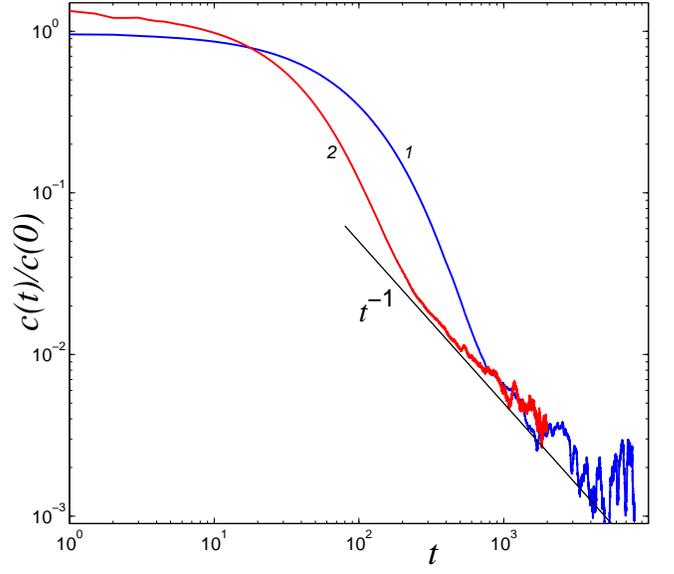}
\caption{
The power-law decay of the correlation function $c(t)$ for the chain with the combined
potentials (\ref{f22}) (curve 1, $b=2.5$, $T=1$) and (\ref{f23})
(curve 2, $T=0.5$). The straight line corresponds to the power law $t^{-1}$.
}
\label{fig18}
\end{figure}

The sum of the Toda and the parabolic potentials has the following form:
\begin{equation}
V(r)=\frac12b^{-2}[\exp(-br)+br-1]+\frac14r^2, \label{f22}
\end{equation}
where the parameter $b>0$, $r=\rho-1$ is the relative displacement.
Simulations show that for $b=2.5$ and $T=1$, the thermal conductivity of the chain $\kappa(N)$
grows as $\ln(N)$ --- see Fig.~\ref{fig17}.
Such an increase is consistent with the observed behavior of the correlation function: the latter
decays slower than $t^{-1}$, see Fig.~\ref{fig18}.
Thus, the chain with the combined potential (\ref{f22}) has an infinite thermal conductivity.

The sum of the parabolic potential and the hyperbolic potential with a hard core (\ref{f12})
has the form of
\begin{equation}
V(r)=\frac14 r^2/(r+1)+\frac14r^2. \label{f23}
\end{equation}
Modeling of the heat transfer in a chain with the combined potential (\ref{f23}) shows
that at $T=0.5$ the thermal conductivity $\kappa(N)$ grows as $\ln(N)$, see Fig.~\ref{fig17}.
Such growth is consistent with the time dependence of the correlation function, which decays
slower than $t^{-1}$, see Fig.~\ref{fig18}.

The sum of the parabolic and Morse potentials has the form of
\begin{equation}
V(r)=\frac12\varepsilon[\exp(-\beta r)-1]^2+\frac14r^2, \label{f24}
\end{equation}
where the parameter $\beta=1/\sqrt{2\varepsilon}$, $\varepsilon=1/72$.
MD simulation of the heat transfer at temperatures $T=0.05$ and $T=0.5$ shows that the correlation
function $c(t)$ for the chain with the pair potential (\ref{f24}) decays as a power function of time,
$\propto t^{-0.9}$ at $t\rightarrow\infty$. Because of this, the thermal conductivity of such
chain also diverges in the limit of $N\rightarrow\infty$.

The non-smooth vibro-impact potential
\begin{equation}
V(r)=\frac12r^2,~\mbox{for}~r>-1,~\mbox{and}~V(r)=\infty~\mbox{for}~r\le -1.
\label{f25}
\end{equation}
can be approximated by a smooth combined potential:
\begin{equation}
V(r)=\frac12 r^2+0.001/(r+1)^2,
\label{f26}
\end{equation}
see Fig.~\ref{fig14}.

MD modeling of the heat transfer in the chain with a combined asymmetric pair potential (\ref{f26})
shows that at $T=1$ the correlation function $c(t)$ also decays as a power function, as $t^{-0.9}$
for $t\rightarrow\infty$. This allows us to conclude that the chain with the vibro-impact
interaction potential (\ref{f25}) has an infinite thermal conductivity.
Thus, the chains with the combined pair interaction potentials, with a parabolic potential as a component,
always are characterized by anomalous heat transport.
The form of the potential (\ref{f25}), its parabolicity for small relative displacements,
allows one to conclude that the anomalous thermal transport is related here with the long mean
free path of small-amplitude phonons.

Analyzing all the simulation results, we can conclude that the asymmetry of the pair potential
only does not guarantee the convergence of the thermal conductivity. The chains with the asymmetric
potential FPU pair potential (\ref{f7}) and with the asymmetric hyperbolic potential (\ref{f11})
have infinite conductivity. The asymptotic behavior of the interaction potential both in
approaching and separating the particles is very important for the convergence of the thermal
conductivity in the thermodynamic limit. If the interaction potential has a hard core (interaction
energy between two particles tends to infinity when approaching their centers) and the
interaction grows no faster than the distance between the particles in separating them, the chain
with such potential will have a finite thermal conductivity. The single-well potentials
(\ref{f8}), (\ref{f12}) and purely repulsive potentials (\ref{f14}), (\ref{f15}) belong to such
class of pair potentials. Morse potential (\ref{f9}) and the short-range potential (\ref{f18n})
do not have hard core, but the chains with these potentials also have  finite thermal conductivity.
The finite binding energy of the pair potential can result in local bond stretching which strongly
scatter phonons. On the other hand, the rapid growth in particles interaction energy in
approaching their centers can result in energy scattering in many-particle collisions. The first
and second scenarios are responsible for the convergence of the thermal conductivity at low
and high temperatures, respectively.

\section{Conclusions}

Our molecular-dynamics simulations show that the one-dimensional chains with unlimited
asymmetric pair potentials can have finite thermal conductivity.
We show that the thermal conductivity converges in the chains with the  Lennard-Jones (\ref{f8})
and  Morse (\ref{f9}) potentials, and in the chain with the hyperbolic potential (\ref{f12}).
The asymmetric potentials, which lead to the convergence of the thermal conductivity, are
characterized by the presence of either a final binding energy, where one branch of the potential
is a limited function [Lennard-Jones, Morse, short-potential (\ref{f18n})], or a hard core
potential [potentials (\ref{f12}), (\ref{f14}), (\ref{f15})].

In the chains  with fixed ends, the  purely repulsive potentials (\ref{f14}), (\ref{f15}),
(\ref{f16}) can be reduced to a single-well asymmetric potentials with an asymptotic  linear branch.
At high temperatures, a chain with such potential may be considered
as a one-dimensional gas of particles, interacting through their collisions.
An example of a chain with a short-range repulsive potential (\ref{f18n}) shows
that the finite conductivity of the "gas" \ is related with the energy dissipation during
many-particle collisions. Therefore the chain with repulsive potentials also have a finite conductivity.
The exception is the potential with exponential repulsion (\ref{f17}), which can be reduced to the
Toda potential (\ref{f20}). Toda lattice is completely integrable
system in which the dynamics can always be described by means of non-interacting
elementary excitations, and therefore it has an infinite thermal conductivity.

In the combined potentials (\ref{f22}), (\ref{f23}), (\ref{f24}) and (\ref{f26}), the parabolic potential,
which stabilizes the bonds, is present. The local strong bond stretching is not possible also in
the chain with the FPU potential (\ref{f7}) and the chain with the vibro-impact potential (\ref{f25}).
There is no anomalous scattering of phonons at the local stretching at low temperatures, and
there is no one-dimensional "gas" \ of particles, interacting through their collisions  at high
temperatures. Therefore, the thermal conductivity is divergent  in the thermodynamic limit.
But the divergence can be very slow and can appear in large lengths only.
That is why it is stated in Refs. \cite{ZZWZ12,CZWZ12} that the thermal conductivity can converge
in the chains with such asymmetric pair potentials. More detailed modeling allows us to
unambiguously conclude that he thermal conductivity diverges in such lattices. Therefore
the chain with the asymmetric FPU potential always has an infinite thermal conductivity.

Note that the Morse potentials are commonly used in molecular dynamics to describe the stiff
valence bonds, and the Lennard-Jones and Coulomb potentials to describe the soft non-valence bonds.
Deformation of the valence and torsion angles is described by the limited angular periodic potentials.
We have shown that the chain with such potentials have a finite thermal conductivity.
Therefore, it is expected that the quasi-one-dimensional molecular systems with such potentials
must have a finite thermal conductivity, but the convergence can occur at very large lengths.
So far these lengths were not reached in MD simulations of carbon nanotubes \cite{SHK09,SavYAKCant12}
and nanoribbons \cite{YAKSav09,SKH10}, and which makes the impression that  the nanotubes
and nanoribbons have infinite thermal conductivity. On the other hand
the convergence of the thermal conductivity was obtained for more soft quasi-one-dimensional molecular
structure, the double helix of DNA \cite{SMKMO}.

Thus, our numerical simulations show that the chains with unlimited
asymmetric potentials that allow the possibility of strong bond stretching
(Lennard-Jones, Coulomb and Morse potentials) are characterized by a finite thermal conductivity.
The convergence of the thermal conductivity is due to scattering of phonons by strongly stretched
bonds at low temperatures, and in result of many-particle collisions at high temperatures.
On the other hand, if the pair interactions limit strong local fluctuations and do not allow for
strong bond stretching, the chain will have an anomalous thermal conductivity. The thermal
conductivity diverges in a thermodynamic limit in a chain with asymmetric FPU potential and in
a chain with any combined potential, which has a parabolic potential as a component.
We can conclude from our simulations that the thermal conductivity of the chain will
diverge if the energy of pair particle interaction grows not  slower than the square
distance in separating the particles.

\section*{Acknowledgements}

A. V. S. thanks the Joint Supercomputer Center of the Russian Academy of Sciences 
for the use of computer facilities.
Yu. A. K. thanks the NANOTHERM project of the European Commission for the financial support
through the Grant Agreement No. 318117,  and the Spanish Ministry of Economy and Competitivity
for the financial support through Grant No. CSD2010-0044. Yu.A.K. acknowledges 
the Ecole Centrale Paris for hospitality.

\end{document}